# An Optics-Based Approach to Thermal Management of Photovoltaics: Selective-Spectral and Radiative Cooling

Xingshu Sun,[1] Timothy J Silverman,[2] Zhiguang Zhou,[1] Mohammad Ryyan Khan,[1] Peter Bermel,[1] and Muhammad Ashraful Alam[1]

[1]Purdue University, West Lafayette, IN, 47907, USA

[2]National Renewable Energy Laboratory, Golden, Colorado, 80401, USA

*Abstract* - For commercial one-sun solar modules, up to 80% of the incoming sunlight may be dissipated as heat, potentially raising the temperature 20°C–30°C higher than the ambient. In the long term, extreme self-heating erodes efficiency and shortens lifetime, thereby dramatically reducing the total energy output. Therefore, it is critically important to develop effective and practical (and preferably passive) cooling methods to reduce operating temperature of PV modules. In this paper, we explore two fundamental (but often overlooked) origins of PV self-heating, namely, sub-bandgap absorption and imperfect thermal radiation. The analysis suggests that we redesign the optical properties of the solar module to eliminate parasitic absorption (*selective-spectral cooling*) and enhance thermal emission (*radiative cooling*). Comprehensive opto-electro-thermal simulation shows that the proposed techniques would cool one-sun terrestrial solar modules up to 10°C. This self-cooling would substantially extend the lifetime for solar modules, with corresponding increase in energy yields and reduced levelised cost of electricity (LCOE).

## I. INTRODUCTION

A typical solar module converts ~20% of the incoming sunlight into electricity. Therefore, up to ~80% of the sunlight may dissipate as heat in the module, causing undesired self-heating as well as performance degradation [1], [2]. Depending on the environment, the average temperature of a solar module can be 20°C–40°C higher than the ambient. The self-heating of PV modules reduces both short-term and long-term power outputs. In the short term, the efficiencies of different PV technologies decrease with temperature, e.g., the efficiency of crystalline Si modules drops by ~0.45% for every 1°C increase in temperature. In the long term, the reliability of modules suffers from thermally activated degradations, such as contact corrosion and polymer degradation, which accelerate at higher temperatures. A recent survey in India has shown that solar modules in hot climates degrade at ~1.5 %/year, eight times faster than the ones installed in cold climates (~0.2 %/year) [2]. The module lifespan was less than 15 years in hot environments, far below the 25-year standard solar panel warranty. As a result, it is important to develop effective cooling schemes to improve both the short-term and the long-term energy yields.

There are several active and passive cooling schemes already in use. These include evaporative and fin cooling [3], liquid submerged PV [4], heat pipe-based system [5], and so on [6]. *These methods cool the panels already heated by the sunlight. A scheme designed to 'prevent' or suppress self-heating could be far more effective*. Modification of the module configuration based on the fundamental physics of self-heating of PV may create a simpler, yet more effective cooling for modules.

In this context, a recent proposal involving radiative cooling of solar cells has drawn much attention [7]–[10]. However, the implication of radiative cooling for practical PV modules is not clear. For instance, Ref. [7], [8] used fused-silica as the starting point for comparison, yet fused-silica is an inferior thermal emitter compared to the commercial coverglass used in PV modules [9]. The role of electricity output of a practical solar module in determining the module temperature was also not accounted for (e.g., a slab of Si wafer instead of a solar cell was assumed in [8], [9] and ideal solar cells at the Shockley-Queisser limit are assumed in [10]). Moreover, thermal radiation from the back side (backsheet) of solar modules was neglected in [7]–[10]. As a result, it has been difficult to ascertain the effectiveness of radiative cooling on commercial PV modules.

In this paper, we explore experimentally the physical origins of PV self-heating for a variety of solar technologies (e.g., Si, CIGS). A large fraction of elevated PV module temperature can be attributed to parasitic sub-bandgap (sub-BG) absorption as well as imperfect thermal radiation to the surroundings. Therefore, we propose to implement a sub-BG optical filter (*selective-spectral cooling*) to eliminate the parasitic absorption, and modify the top and bottom surfaces (*radiative cooling*) to enhance thermal emission. The cooling design is validated by our self-consistent opto-electro-thermal coupled simulation. We predict substantial temperature reduction for different PV materials. For example, we expect ~6 °C and ~10 °C temperature reductions in Si and CdTe solar modules, respectively.

The paper is organized as follows. In Sec. II, we discuss the balance of energy fluxes in solar modules by introducing our opto-electro-thermal coupled framework. The underlying physics of PV self-heating is explored in Sec. III, and the corresponding optics-based cooling methods (i.e., selective-spectral and radiative cooling) are presented in Sec. IV. The cooling effectiveness is investigated in Sec. V, and its implication on both short-term and long-term energy yields is discussed in Sec. VI. Finally, we conclude the paper in Sec. VII.

## II. OPTO-ELECTO-THERMAL COUPLED FRAMWORK

**Energy Fluxes.** A *terrestrial* PV module is subject to the following energy fluxes, see Fig. 1: 1) the absorbed solar irradiance, $P_{Sun}$, determined by the solar spectrum (e.g., AM1.5) as well as the absorptivity of the PV module; 2) the sky cooling, $P_{Sky}$, through radiative energy exchange with the atmosphere from the side facing the sky; 3) similarly, cooling due to energy transfer to the ground, $P_{Ground}$ from the backside; 4) convective cooling by air at the top and bottom surfaces and conductive heat transfer through the aluminum frames, $P_{Conv(d),top/bottom}$; 5) most importantly, the output power delivered by PV modules to the external load, $P_{Out}$.

**Energy-balanced equations.** For a thermodynamic system in the steady state, the incoming and outgoing energy fluxes should balance out to reach equilibrium; namely,

$$P_{Sun} = P_{Sky} + P_{Ground} + P_{Conv(d),top} + P_{Conv(d),bottom} + P_{Out}, \quad (1)$$

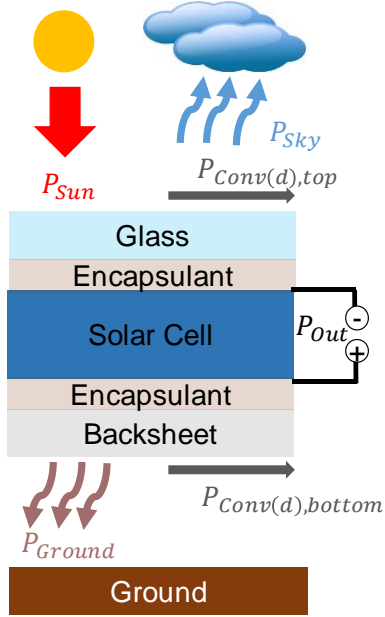

Fig. 1 Schematic of a terrestrial PV module, where we have identified the incoming and outgoing energy fluxes. Eq. (1) summarizes the energy-balance equation for the solar module.

for terrestrial solar modules. Note that each energy flux in (1) are determined by the thermal state and optical properties of the PV modules as well as the outside environment. So one must solve (1) opto-electro-thermally and self-consistently to calculate the steady-state temperature of PV modules. For instance, o*ptically,* we calculate $P_{Sun}$ by integrating the measured absorptivity and the solar spectrum. *Thermally,* $P_{Sky}$ depends on the temperature of PV modules, $T_{PV}$, and the ambient temperature, $T_A$, as well as the emissivity of PV modules and atmospheric transmittance in the infrared (IR) region. *Electrically,* the output power, $P_{Out}$, is temperature-dependent and varies among different PV technologies. Finally, the calculated temperature at equilibrium must give energy fluxes that satisfy (1). A summary of equations to calculate the energy fluxes are specified in the Appendix. Unlike the empirical approaches in [11], [12], the opto-electro-thermal simulation framework in this work can physically calculate operating temperature of modules with different solar absorbers (e.g. Si, CIGS) and various environment conditions without any fitting parameters.

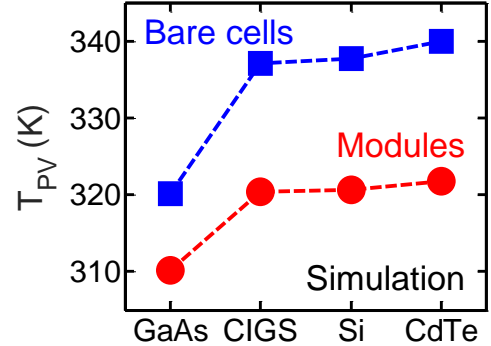

Fig. 2 The outdoor operating temperature of bare cells (blue squares) and encapsulated modules (red circles) of GaAs, CIGS, Si, and CdTe.

**Benchmark against experiments.** Fig. 2 shows the temperature calculated by our opto-electro-thermal framework for different PV technologies under the same environment conditions (i.e. the wind speed is ~0.5 m/s giving an effective convective coefficient h=10 W/(K.m$^2$) [13]; conductive heat transfer only at the module edges through metal frames is neglected; the atmospheric transmittance data is in Fig. 4; the ambient temperature $T_A$ and solar irradiance are 300 K and 1000 W/m$^2$, respectively;). There are also two interesting observations from the simulated data: 1) the operating temperature varies among different PV technologies. Specifically, GaAs modules operate at much lower temperature (~310 K) compared to the others. Remarkably, our simulation anticipates the following two trends observed in the outdoor tests: (a) commercial GaAs modules operates at lower temperature (~ 10 K) compared to Si-based solar cells [14], and (b) an encapsulated module operates at lower temperature (10-20 K) compared to a bare cell without coverglass [8]. Indeed, these two observations can be attributed to two important self-heating mechanisms in photovoltaics: a) parasitic sub-BG absorption and b) imperfect thermal radiation, which will be discussed in detail in Sec. III.

## III. PHYSICAL ORIGINS OF SELF-HEATING

### A. Parasitic sub-BG absorption

The solar irradiance consists of photons ranging from the ultraviolet spectrum (~4 eV) to near-IR region (~0.5 eV). In general, however, only photons with energy above the bandgap excite electron-hole pairs in a semiconductor to produce electricity. For Si or CIGS solar cells ($E_G \approx 1.1$ eV), the above-bandgap spectrum accounts for ~84% of the incident solar irradiance. A module with ~18% efficiency converts part of the above-bandgap solar energy into electricity, the rest is converted to heat through carrier recombination, thermalization, and entropy generation [15]. One way to lower heat generation from above-bandgap photons is to increase the intrinsic solar cell efficiency (by multi-junction design [16], etc.), which is not discussed in this paper because we wish to focus on single-junction cells. On the other hand, for Si and CIGS, ~16% of the sunlight consists of photons with energy below the bandgap. Ideally, the sub-BG photons will not be absorbed by solar cells, rather it should be reflected back by the back metal.

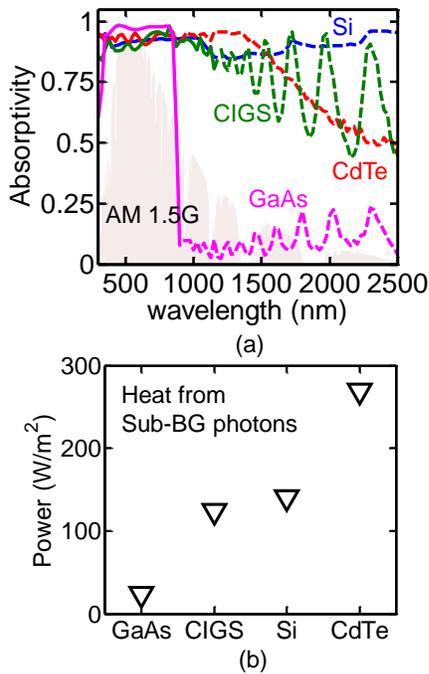

Fig. 3 (a) Measured absorptivity for different solar absorber materials vs. photon wavelength (solid lines: above bandgap photons; dashed lines: below bandgap photons). The pink area is AM1.5G spectrum. (b) Heat from sub-BG photons for different technologies.

We have measured the absorptivity profile of four different samples, with particular emphasis on the sub-BG spectrum. The optical measurements were performed using an Agilent-Cary 5000 spectrophotometer (with an integrating sphere) [17] at the National Renewable Energy Laboratory (NREL). The Si sample was a commercial solar module from [14], and GaAs [18], CIGS [19] and CdTe [20] samples were fabricated at NREL lab. All the cells (except CIGS) had anti-reflection coating. The cell-level measurement, however, may underestimate the parasitic absorption slightly, because ~3% of sunlight is absorbed by in the encapsulation layers of a practical module structure [21]. Otherwise, the absorptivity profile of a module is essentially the same as that of ARC-coated bare cell, an assertion validated by our numerical modeling (not shown).

*Our measurements of different PV technologies, however, show various degrees of sub-BG absorption (dashed lines in Fig. 3(a))*. Specifically, Si, CIGS, and CdTe show high sub-BG absorption, while most of the below-bandgap photons are reflected in GaAs. The parasitic absorption may be variously attributed to absorbing back metal reflector, the Urbach tail, as well as free carrier absorption by highly-doped layers (emitter and back surface field in Si or window and buffer layers in CIGS and CdTe) [22]–[24]. Consequently, a large fraction of the sunlight, which consists of the sub-BG photons, now heats the solar module, see Fig. 3(b).

Among these technologies, GaAs is almost immune to sub-BG absorption possibly due to the high-quality metal mirror (gold) and reduced free carrier absorption. The magnitude of sub-BG absorption is similar between CIGS and Si (~12 % of the solar irradiance). Interestingly, CdTe has the largest parasitic absorption (~30 %) due to its larger bandgap (~1.5 eV) and strong absorptivity in the sub-BG spectrum. The consequence of sub-BG absorption among different technologies is reflected in Fig. 2, i.e., GaAs and CdTe operate at the lowest and highest temperatures, respectively. Obviously, the sub-BG absorption is not an intrinsic property of a cell technology (it can be reduced by modifying cell design, for example), therefore, the purpose of the discussion above is to highlight the importance of sub-BG absorption in determining the operating temperature of solar modules. Consequently, it is desired to eliminate the sub-BG absorption, which contributes substantially to self-heating, but not to the output power. In Sec. III, we will propose to redesign solar modules optically such that sub-BG photons are not absorbed. Next, however, we will discuss another source of self-heating, namely, imperfect thermal radiation of dissipated heat.

*B.    Imperfect thermal radiation*

**Thermal radiation for cooling.** Another important factor dictating operating temperature of PV ($T_{PV}$) is the constant exchange of energy between the module and the surroundings through thermal radiation. Solar modules at outdoors receive thermal radiation from the sky and the ground; meanwhile, the top (glass) and bottom (polymer backsheet) layers of PV modules radiate to the sky and the ground, respectively. Given that the daytime module temperature is higher than the ambient, the net energy exchange from modules to surroundings is positive. Therefore, the ambient environment cools modules through thermal radiation with a spectrum peaking in the IR wavelengths. Without the cover-glass, however, solar absorbers can display very low emissivity in the IR spectrum, see Fig. 4. Hence, the amount of emitted thermal radiation is substantially suppressed for a bare solar cell, resulting in much higher temperature, as shown in Fig. 2. As a result, even though cell-level measurements are usually conducted indoors with heat sinks to maintain constant temperature, one must be careful to interpret the results from outdoor cell-level measurements.

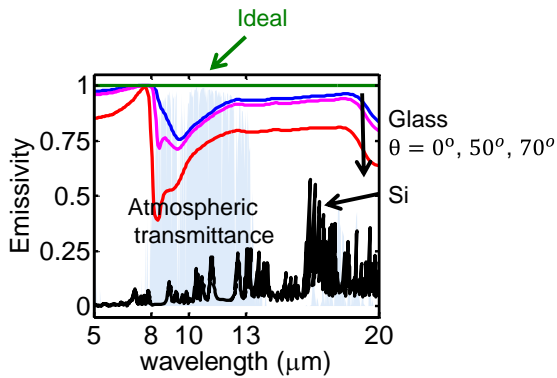

Fig. 4  Simulated emissivity profile of glass at different incident angles θ using S$^4$ [41]. The (n, k) data was obtained from [42]. The emissivity of Si is obtained from [7]. The ideal emissivity for radiative cooling is also shown here as green line. The blue area is the atmospheric transmittance in the zenith direction calculated by ATRAN [43] for New Delhi in spring with perceptible water vapor (PWV) = 18 mm.

**Imperfect thermal radiation.** Despite the fact that glass and backsheet are already highly emissive in the IR region, they are still not perfect. The emissivity of glass is calculated in Fig. 4, which shows a drop of the emissivity in the atmospheric transmission window (blue shaded area). The window corresponds to the wavelength range (8 $\mu$m –13 $\mu$m) where the atmosphere is transparent (high transmittance) to thermal emission. It is also noteworthy that the wavelengths of peak thermal radiation from many terrestrial objects exactly match the "transparent" window. In other words, objects on Earth can exchange a large amount of energy with the cold troposphere (usually 50 K lower than the ambient temperature at sea level) through these wavelengths. Hence, any dip of the emissivity between 8 μm and 13 μm can lower the cooling power of a thermal emitter. Also, the emissivity of glass at higher angles reduces rapidly beyond 50º, see Fig. 4. Since thermal radiation is hemispheric (integrated with angles from 0º to 90º), the angle-dependent emissivity of glass reduces the thermal radiation from solar modules compared to an ideal emitter. Overall, the calculated average emissivity (hemispherical emissivity) is 0.82 very close to the commercial solar glass ($\bar{\varepsilon}$ = 0.84) [25], while commercial PVF backsheet has $\bar{\varepsilon} \approx$ 0.85 [26], i.e., both have room for improvements. Therefore, it is desirable to re-engineer the top and bottom surfaces of solar modules to enhance thermal radiation for cooling, as we will discuss in Sec. IV.

## IV. OPTICS-BASED COOLING METHODS

Thermodynamics dictate that modules must self-heat, but our focus is on avoidable temperature rise due to a) strong sub-BG absorption, b) inadequate thermal radiation. To mitigate this parasitic self-heating, we propose two optics-based cooling methods, namely, selective-spectral cooling and radiative cooling. We will briefly discuss the practical implementation or the economic viability of these cooling methods in the Sec. VI; for now, we focus on the effectiveness of the ideal designs in reducing the module temperature.

### A.  Selective-spectral cooling

Ideally, since the sub-BG photons do not contribute to the electricity output, they should be reflected by the cells or modules. Instead, our measurements in Fig. 3 show a large fraction of sub-BG photons are absorbed by the cell (e.g., ~300 W/m$^2$ for CdTe), which in turn heats up the solar module. Note that the parasitic absorption is related to the intrinsic material properties of PV modules (e.g., free carrier absorption, reflection loss), and it is not trivial to eliminate the parasitic absorption by improving absorber materials. An alternative approach may involve selective reflection the sub-bandgap photons *before* they enter the solar absorber by implementing optical filters or selective mirrors, see Fig. 5.

Ideally, the optical filter in Fig. 5(a) should be a short-pass filter, which only allows photons above $E_G$ to pass and reflect the rest. Such a filter can be realized using quarter-wave stacks [27]. It is important that the filter does not interfere with sky-cooling, therefore, the optical filter should be inserted in between coverglass and polymer encapsulant. The filter can also be engineered to reflect the high-energy ultraviolet photons, which does not contribute efficiently to carrier generation, but cause polymer yellowing and encapsulation delamination [28] [29]. We, however, will not study or optimize for the latter.

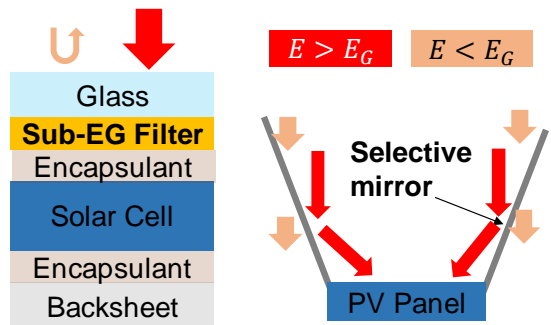

(a) Optical Filter     (b) Selective-Concentration

Fig. 5  Possible implementations of selective-spectral cooling by using a reflective optical filter or wavelength-selective mirror reflector for LCPV.

Selective-spectral cooling can also be particularly interesting for low-concentration PV (LCPV) applications, where the heat from sub-BG photons scales with concentration factor, but without the benefit of active cooling. For LCPV, side mirrors are used to concentrate sunlight onto PV modules. The widely-used metal-coated mirrors, however, have the disadvantage of reflecting the near-IR sunlight, which is dissipated as heat in PV modules. One potential improvement is to adopt wavelength-selective mirror using nanophotonics [30] or IR transmissive polymeric films [31], [32] such that only the useful photons are directed to solar modules and the rest just pass through the mirror, see Fig. 5(b). Self-heating due to sub-BG photons is therefore reduced.

*B. Radiative cooling*

As discussed in Sec. III, the top (glass) and bottom (polymer backsheet) layers of PV modules are not ideal in terms of emitting IR thermal radiation to the atmosphere and the ground. Hence, we propose to add radiative cooler layers to enhance thermal radiation from PV modules to the surroundings. The radiative cooler on top of the glass should have the ideal emissivity profile in Fig. 4 for maximum thermal emission but must be transparent below 2.5 $\mu$m wavelength for solar irradiance. For objects at temperatures close to 300 K, thermal radiation shorter than 2.5 $\mu$m wavelength is negligible (~ 0.02 W/m$^2$ at 340 K). Hence, the transparency shorter than 2.5 $\mu$m does not sacrifice much radiative cooling power. In principle, such spectral response can be achieved using a nanophotonic crystal [8], [33]. An ideal blackbody can be used on the back surface to maximize thermal radiation exchange with the ground, but one can still use the radiative cooler for the back layer, since its performance is very close to a blackbody for IR radiation near 300 K. Note that those selective emitters which restrain thermal radiation between 8 $\mu$m and 13 $\mu$m in [34], [35] are not suitable for cooling solar modules. The hemispherical emissivity of such emitters ($\bar{\varepsilon}$ = 0.32) is far below that of glass ($\bar{\varepsilon}$ = 0.82), and actually would lead to higher temperature of solar modules. Those designs are only of great interest for cooling below the ambient, which solar modules illuminated under sunlight cannot achieve because solar irradiance (1000 W/m$^2$) is greater than thermal radiation of objects at ~300 K.

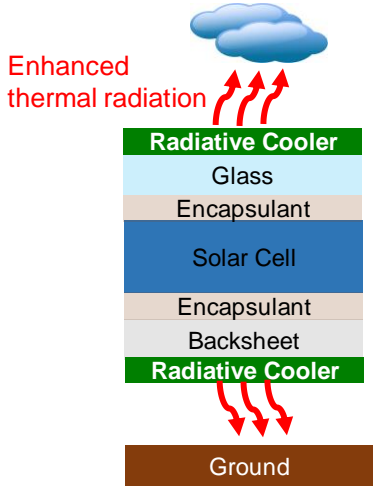

Fig. 6   Schematic of a solar module with enhanced radiative cooling.

## V. RESULTS

An interesting question is how much temperature reduction can be obtained by the two aforementioned cooling methods. To answer this question, we explored the cooling effects using our opto-electro-thermal coupled modeling framework to simulate the one-sun solar module temperatures with and without cooling. The simulation assumes ideal scenarios of the cooling methods (*i.e.*, ideal filter with cutoff at $E_G$ for selective-spectral cooling and unity IR emissivity for radiative cooling), which reveals the theoretical maximum reduction of temperature.

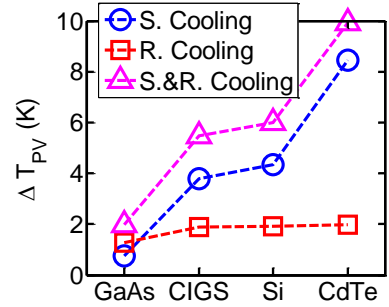

Fig. 7   Temperature reduction (with respect to module temperatures in Fig. 2) using different cooling methods (S. Cooling: selective-spectral cooling; R. Cooing: radiative cooling; S.&R. Cooling: selective-spectral cooling and radiative cooling combined) for different technologies.

Fig 7 illustrates the temperature reduction ($\Delta T_{PV}$) using the cooling schemes, compared to the module temperatures in Fig. 2. One important observation is that the selective-spectral cooling method can reduce module temperatures by ~4 K for CIGS and Si and ~8 K for CdTe, but only ~0.5 K for GaAs. This is because most of the sub-BG photons are already reflected in GaAs and further filtering these photons do not provide efficient cooling. Perfect radiative cooling provides limited cooling benefits (~1 K to 2 K reduction) compared to glass covered modules for all technologies, which agrees with the calculation in [9]. The results indicate that replacing glass ($\bar{\varepsilon}$ = 0.82) and PVF backsheets ($\bar{\varepsilon}$ = 0.85) with ideal thermal emitters does not result in a large decrease in the temperatures of conventional terrestrial PV modules. By applying both cooling schemes simultaneously, one can achieve a superposed temperature reduction. The additive cooling is understandable since these two cooling methods address different sources of PV self-heating, namely, parasitic sub-BG absorption and imperfect thermal radiation.

## VI. DISCUSSION

**Environmental factors dictate self-cooling.** So far, we have calculated $T_{PV}$ by assuming the ambient temperature $T_A = 300$ K, an effective convective coefficient $h = 10$ W/(K.m$^2$) (~0.5 m/s wind speed), and the atmospheric transmittance for New Delhi in spring (Fig. 4). The remaining question is how environmental factors change the cooling effect. **First**, as $h$ increases in a windier condition, more of the heat is lost through convection. Hence, the effectiveness of spectral and radiative cooling (reflected in absolute $\Delta T_{PV}$, see Fig. 8(a)) is reduced at higher wind speeds (higher $h$), because the excess heat is carried away by convection. Since wind speed depends on the season and the geographical location (e.g., average monthly wind-speed in New Delhi is around 4.2 m/s and 0.8 m/s in June and October, respectively [36]), the overall effectiveness of the self-cooling strategies must be evaluated carefully for a solar farm installed in a given geographical location. **Second**, at a fixed wind speed, radiative cooling is more effective in a hotter climate as shown in Fig. 8 (b), because thermal radiation power scales with temperature as $P \sim T_{PV}^4$. On the other hand, intrinsic power loss (e.g., carrier recombination) increases with temperature, leading to more heat dumped from the above-bandgap irradiance. Hence, reflecting the heat power from sub-BG photons, i.e., selective-spectral cooling, is slightly less effective with increasing $T_A$, as shown in Fig. 8 (b). Even though selective-spectral and radiative cooling show different trends with the ambient temperature, the cooling gain by integrating these cooling methods is almost independent of $T_A$. **Third**, the degree of cooling depends on the illumination intensity, see Fig. 8 (c). Since the heat dissipated in the module is reduced at lower illumination, the relative efficiency improvement by the proposed cooling techniques is also suppressed at lower illumination. **Finally**, the presence of water vapor and $CO_2$ reduces the transmittance between 8 μm and 13 μm of the atmosphere, directly suppressing thermal radiation from the glass encapsulation to the outer space [34]. Consequently, radiative cooling is expected to be less useful in humid and cloudy climates.

**Benefits of Cooling.** We have demonstrated temperature reduction of the cooling methods on different PV technologies. The next obvious question is: how much energy yield gain can be achieved by cooling PV modules? For Si solar modules in terrestrial environments with an average ambient temperature of 300 K and wind speed of 0.5 m/s, the highest temperature reduction by applying the cooling methods is 6 K for Si commercial modules. Given the typical temperature coefficient $\beta \approx -0.45$ %/K of Si, 6 K can provide 2.7 % improvement to the short-term electricity output, corresponding to 0.5 % absolute increase in the efficiency of Si solar modules. Hence, the proposed cooling methods offer an alternative way to improve the efficiencies without changing the intrinsic material properties of the solar cells.

What about long-term energy gain due to self-cooling? Most degradation processes, such as moisture ingress and potential-induced degradation, are thermally activated; according to an Arrhenius relationship, the time to failure of solar modules is proportional to $\exp(-E_A/k_B T)$, where $E_A$ is the effective activation energy and $k_B$ is the Boltzmann constant. Using the calibrated average activation energy, $E_A = 0.89$ eV, accounting for a variety of degradation mechanisms (e.g., corrosion of interconnect, EVA yellowing, potential-induced degradation) [37] and the empirical equation for lifetime from [38], *6 K reduction in average operating temperature can delay PV module failure due to thermally activated degradation by up to ~85%*. As a result, selective-spectral and radiative cooling can offer significant reliability improvements and greatly reduce the levelised cost of electricity (LCOE).

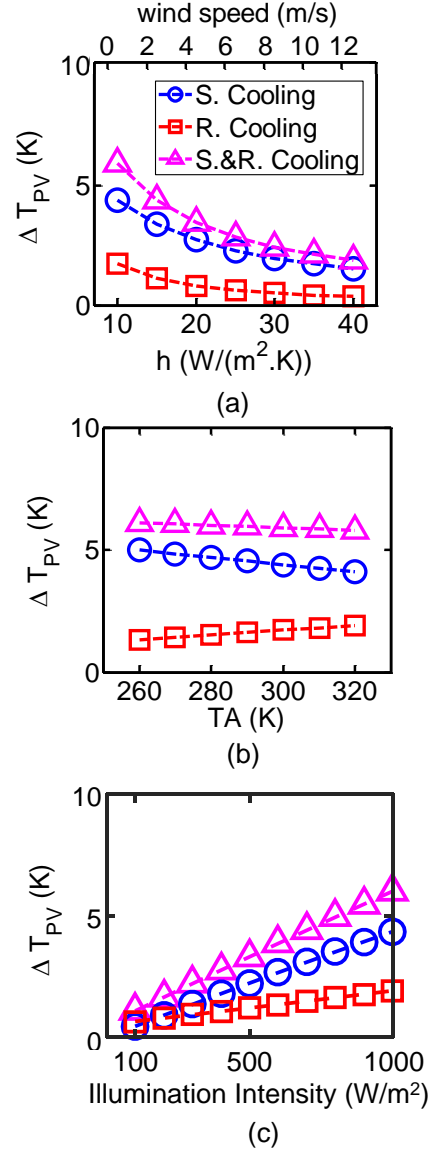

Fig. 8 Temperature reduction of conventional Si modules as a function of (a) convective coefficient/wind speed, (b) the ambient temperature, and (c) the illumination intensity. The default environment parameters for simulation are $T_A = 300$ K, $h = 10$ (W/(m$^2$.K)), and illumination = 1000 W/m$^2$. The atmospheric transmittance is taken from Fig. 4.

**Environmental Factor.** So far, the calculation of short- and long-term energy gains due to cooling has assumed a constant average ambient temperature of 300 K, solar irradiance of 1000 W/m$^2$, and wind speed of 0.5 m/s ($h = 10$ W/(K.m$^2$)). In practice, the increase of energy yield of a PV module over the course of an entire year depends on the local environment (e.g., illumination, wind speed, relative humidity, and ambient temperature). For example, the effectiveness of selective-spectral and radiative cooling is reduced at locations with high wind speed, because the module temperature is already low, and the additional benefits of selective-spectral/radiative cooling is relatively small. In addition, solar modules installed in environments with higher humidity and higher ambient temperature degrade substantially faster; hence, cooling the solar modules will significantly enhance the reliability and boost integrated energy yield. Hence, one must properly account for the geographic and temporal variation of the environmental factors to accurately predict all the incremental electricity yields by adopting the approaches discussed in this paper.

**Selective-spectral vs. Radiative Cooling.** Integrating selective-spectral and radiative cooling provides the most cooling advantages for solar modules, but one also needs to consider the feasibility and cost in practice. Zhu [8] has demonstrated experimentally the use of a photonic crystal (PhC) structure to improve the hemispherical emissivity for radiative cooling but the emissivity still drops substantially at higher incidence angles (Fig. 5(b) in [8]) and the hemispherical emissivity is estimated to be around 0.9, still far from unity. The fabrication cost of a nano-photonic structure also makes it an impractical option for large-scale manufacture. Additionally, though Ref. [7] argues that PhC structure can exhibit hydrophobicity and self-cleaning function, the potential soiling issues from the deep air holes in PhC still need to be carefully considered especially in environments lack of rain water. Other high-emissivity coverglass applications have also been explored especially for extraterrestrial PV modules, such as pseudomorphic glass (PMG) [39]. The economic viability of adopting such a glass-technology, especially for large-scale terrestrial solar farms, remains an interesting open question.

On the other hand, selective-spectral cooling in general is more beneficial than radiative cooling, making selective-spectral cooling much more preferable. Optical filters with customized wavelength selectivity are commercially available and may be suitable for large-scale manufacturing. Including additional UV blocking in the filter can further prevent performance degradation from yellowing and delamination of encapsulants [28], [29] . The non-ideal sharpening of the filter which can degrade short circuit current and the tradeoff between cutoff sharpness and pass-band transmissivity must be carefully engineered. It also is important to note that the bandgaps of Si and GaAs decrease with temperature, characterized by the temperature coefficient (-4.73 x $10^{-4}$ eV/K for Si and -5.41 x $10^{-4}$ eV/K for GaAs), which may affect the optimal cutoff wavelength of the filter. The variation of bandgaps, however, is very small (~0.01 eV in the temperature range of interest for one-sun solar modules (300 K to 320 K). For concentrated PV with much higher operating temperature, the cutoff of the filter has to be optimized carefully to account for the temperature-dependence of bandgap. Alternative ways for selective-spectral cooling include de-texturing the front layer or nitridizing the back surface field in Si modules, both of which have been demonstrated experimentally [22], [40]. Hence, selective-spectral cooling can be more advantageous than radiative cooling for conventional solar modules, unless cost-friendly cover materials with high IR emissivity and solar transmittance are discovered. However, radiative cooling could be very effective for extraterrestrial solar modules in the absence of air convective cooling. Therefore, for both space and concentrated PV, radiative cooling remains promising to be further explored.

VII. CONCLUSIONS

To summarize, we find that self-heating in PV modules has large components due to parasitic sub-BG absorption and inadequate thermal radiation. These results are confirmed by measurements of different solar technologies (i.e., GaAs, CIGS, Si and CdTe) and outdoor tests in literature [8], [14]. To address these issues, we have proposed to optically redesign solar modules by implementing selective-spectral cooling (i.e., eliminate sub-BG parasitic photon absorption) and radiative cooling (i.e., enhance thermal radiation to the surroundings). Substantial temperature reduction has been demonstrated in different PV technologies based on our self-consistently opto-electro-thermal simulation. Potentially, the temperature reduction can provide 0.5% absolute increase in efficiency and extend the lifetime by 80% for one-sun Si terrestrial solar modules. We also predict that selective cooling is likely to be more cost-competitive as well as more effective than radiative cooling for conventional solar modules, while the prospects of using radiative cooling in concentrated and extraterrestrial PV remain encouraging. The effectiveness of these cooling methods bring new potentials to improve reliability and performance of photovoltaics.

ACKNOWLEDGMENT

This work is made possible through financial support from the National Science Foundation through the NCN-NEEDS program, contract 1227020-EEC and by the Semiconductor Research Corporation, the US-India Partnership to Advance Clean Energy-Research (PACE-R) for the Solar Energy Research Institute for India and the United States (SERIIUS), U.S. Department of Energy under Contract No. DE-AC36-08GO28308 with the National Renewable Energy Laboratory, the Department of Energy under DOE Cooperative Agreement No. DE-EE0004946 (PVMI Bay Area PV Consortium), and the National Science Foundation under Award EEC1454315-CAREER: Thermophotonics for Efficient Harvesting of Waste Heat as Electricity. The authors would like to thank Dr. Dave Albin, Dr. Lorelle Mansfield, Dr. Ingrid Repins, and Dr. Myles Steiner for the measurement data, Rajiv Dubey, Shashwata Chattopadhyay, Raghu Vamsi Chavali, and Dr. Xufeng Wang for helpful discussion, and Prof. Anil Kottantharayil, Prof. Juzer Vasi, and Prof. Mark Lundstrom for kind guidance.

APPENDIX

In this appendix, the equation to calculate each energy flux in (1) is presented. The absorbed sunlight can be written as

$$P_{Sun} = \int_0^\infty d\lambda I_{Sun}(\lambda) \times \varepsilon(\lambda, \theta_{Sun}) \times \cos(\theta_{Sun}), \quad (A1)$$

where $\theta_{sun}$ is the solar incidence angle ($\theta_{Sun} = 0^o$ in this work), $I_{Sun}(\lambda)$ is spectral flux density of the solar spectrum at different wavelengths $\lambda$ and $\varepsilon(\lambda, \theta_{Sun})$ is the absorptivity of solar modules at incidence angle $\theta_{sun}$. For conventional solar modules, $I_{Sun}(\lambda)$ is the AM1.5G spectral density, while AM1.5 D and AM0 spectrums should be used for concentrated and extraterrestrial PV, respectively.

Sky cooling power in (1) for terrestrial modules is

$$P_{Sky}(T_{PV}, T_A) = P_{Rad}(T_{PV}) - P_{Atm}(T_A). \quad (A2)$$

In Eq. (A2), $P_{Rad}(T_{PV})$, the thermal emission power radiated from the glass cover for both terrestrial and extraterrestrial modules can be expressed as

$$P_{Rad}(T_{PV}) = \int d\Omega \cos(\theta) \int_0^\infty d\lambda I_{BB}(T_{PV}, \lambda) \times \varepsilon(\lambda, \Omega). \quad (A3)$$

Here, $\varepsilon(\lambda, \Omega)$ is the angular emissivity of glass; $I_{BB}(T, \lambda) = (2hc^2/\lambda^5)/(\exp(hc/(\lambda k_B T)) - 1)$ where $h$ is the Plank constant, $c$ is the velocity of light, and $k_B$ is the Boltzmann constant; $\int d\Omega = \int_0^{\pi/2} d\theta \sin(\theta) \int_0^{2\pi} d\phi$ is the angular integral over a hemisphere. Similar, $P_{Atm}(T_A)$ which is the thermal radiation from the atmosphere to PV modules can be written as

$$P_{Atm}(T_A) = \int d\Omega \cos(\theta) \int_0^\infty d\lambda I_{BB}(T_A, \lambda) \times \varepsilon(\lambda, \Omega) \times \varepsilon_{Atm}(\lambda, \Omega). \quad (A4)$$

Using Kirchhoff's law and the Beer-Lambert law [34], the angular emissivity of the atmosphere $\varepsilon_{Atm}(\lambda, \Omega)$ can be written as $\varepsilon_{Atm}(\lambda, \Omega) = 1 - t_{Atm}(\lambda)^{1/\cos(\theta)}$, where $t_{Atm}(\lambda)$ is the atmospheric transmittance in the zenith direction in Fig. 4. As pointed out in [34], the downward atmospheric spectrum can be divided into two sub-spectrums: the first one spanning 8-13 $\mu m$, and the second involving the rest of the wavelengths. The 2$^{nd}$ spectrum (outside the 8 -13 $\mu$m wavelength range) is emitted by water vapor and carbon dioxide within the lowest few hundred meters of the sky, at the local ambient temperature $T_A$. In contrast, the '8 - 13 $\mu m$ spectrum' stems from the upper part of the troposphere with $T < T_A$. Hence, the atmosphere has lower spectral emissivity within 8 - 13 $\mu m$ wavelength, see Fig. 3 in [34]. Because the emissivity depends on wavelength, we calculate the atmospheric radiation (see A4) by integrating the Planck's equation (at $T_A$) with the atmospheric emissivity, $\varepsilon_{Atm}(\lambda, \Omega)$, over the entire IR wavelength range.

Since wavelength-dependent emissivity of backsheet is not available, cooling power of thermal radiation exchange between the bottom surface and the ground (Earth) is calculated using the Stefan–Boltzmann law as

$$P_{Ground}(T_{PV}, T_A) = \sigma \varepsilon F (T_{PV}^4 - T_A^4), \quad (A5)$$

where $\varepsilon$ is the hemispherical emissivity of the back surface, $F$ is the view factor and $\sigma$ is the Stefan–Boltzmann constant. The ground temperature (could be slightly lower than $T_A$ in practice) is assumed to be the same as the ambient temperature in this work. The view factor is assumed to be unity for terrestrial (i.e., no tilting) solar modules in this paper.

The convective cooling power is calculated by

$$P_{Conv(d)}(T_{PV}, T_A) = h \times (T_{PV} - T_A), \quad (A6)$$

where $h$ is the effective heat transfer coefficient combing the free and forced convection and conduction. In this paper, the effective heat transfer coefficient, $h$, is set to be same for the top and bottom surfaces of solar panels assuming no tilting.

Finally, the electrical output power $P_{Out}(T_{PV})$ of the PV modules is

$$P_{Out}(T_{PV}) = P_{Out}(300 \text{ K}) \times (1 + \beta \times (T_{PV} - 300 \text{ K})). \quad (A7)$$

Here, for a given PV technology, $P_{Out}(300 \text{ K})$ is the output power at 300 K and $\beta$ is the temperature coefficient, which is negative for most solar technologies.

Coupling (A1) to (A7) into (1), one can self-consistently solve the temperature of solar modules under different environmental conditions.